\documentclass[referee]{aa} 
\usepackage{txfonts}
\usepackage{graphicx}
\begin{document}
   \title{A photometric study of the old open clusters
Berkeley~73, Berkeley~75 and Berkeley~25}
\author{
G. Carraro\inst{1,2},
D. Geisler\inst{3},
A. Moitinho\inst{4},
G. Baume\inst{5}, and
R.A.  V\'azquez\inst{5}
}
\offprints{G. Carraro}
\mail{gcarraro@das.uchile.cl}

\institute{
Departamento de Astronom\'ia, Universidad de Chile,
Casilla 36-D, Santiago, Chile
\and
Astronomy Department, Yale University,
P.O. Box 208101, New Haven, CT 06520-8101 , USA
\and 
Universidad de Concepci\'on, Departamento de Fisica,
Casilla 160-C, Concepci\'on, Chile
\and
CAAUL, Observat\'orio Astron\'omico de Lisboa, Tapada da Ajuda,
  1349-018 Lisboa, Portugal
\and
Facultad de Ciencias Astron\'omicas y Geof\'{\i}sicas de la
UNLP, IALP-CONICET, Paseo del Bosque s/n, La Plata, Argentina
            }
\date{Received March 2005; accepted}

\abstract{CCD BVI photometry of the faint open clusters Berkeley~73, 
Berkeley~75 and Berkeley~25 are presented. The two latter 
are previously unstudied clusters to our knowledge. While
Berkeley~73 is found to be of intermediate-age (about 1.5 Gyr old),
Berkeley~75 and Berkeley~25 are two old clusters, with ages
larger than 3.0 Gyr. We provide also estimates of the clusters
size. Very interestingly, all these clusters
turn out to lie  far away from the Galactic Center,
at $R_{GC} \geq$  16 kpc, and quite high onto the Galactic plane,
at $|Z_{\odot}| \geq  1.5$ kpc.
They are therefore important targets to probe the properties of the 
structure of the Galaxy in this direction, where the Canis Major 
over-density has been recently discovered to be located.
}
\titlerunning{Photometry of three old open clusters}
\authorrunning{G. Carraro et al.}

\maketitle
%

\section{Introduction}
The paper belongs to a series aimed at providing homogeneous
photometry of open clusters located in the  third Galactic Quadrant.
The main motivations of the survey are discussed in Moitinho (2001)
and Giorgi et al. (2005). Briefly,  we want to {\it i)} better trace the spiral
structure  and {\it ii)} probe the history of star formation
in this part of the Galaxy.
For these purposes open clusters of different ages
are very useful.
While the youngest clusters are very well known spiral arm tracers, the older
ones are routinely used to probe the chemical evolution
and star formation history of the disk (Carraro et al. 1998,
de la Fuente Marcos \& de la Fuente Marcos 2004).
Moreover very distant open clusters in the Third Galactic Quadrant might help
to probe the stellar population of the recently discovered
Canis Major over-density, or to better understand the warp structure
of the Galactic disk (Momany et al. 2004, Bellazzini et al. 2004). 
In this paper we discuss BVI photometry of three old open clusters,
namely Berkeley~73, Berkeley~75 and Berkeley~25 (= Ruprecht~3) with the aim
to derive their fundamental parameters, in particular age and distance,
which are the basic input for the derivation of the star
formation history in the Galactic disk. 
These clusters until now were unstudied except for Berkeley~73,
whose first Color Magnitude Diagram (CMD) is discussed in 
Ortolani et al. (2005).
The coordinates
of these three clusters are listed in Table~1, and have been
redetermined  by one of us (G.C.) based on inspection of 
Digital Sky Survey (DSS and XDSS) images.
In fact these clusters are very loose and faint, and in many cases
the reported coordinates are not very precise.\\
e notice that the coordinates reported by Ortolani et al. (2005)
for Berkeley~73 are wrong, and refer instead to the cluster vdB-Hagen 73.\\

\noindent
The plan of this paper is as follows.
In Sect.~2 we briefly illustrate observations and data reduction and
in Sect.~3 we describe how we obtain astrometry
In Sect.~4 we derive an estimate of the clusters diameters, and
in Sect.~5 we describe the CMDs
Sect.~6
is then dedicated to derive clusters' fundamental parameters
Finally, in Sect.~7 we draw some
conclusions and suggest further lines
of research.

\begin{table}
\caption{Basic parameters of the clusters under investigation.
Coordinates are for J2000.0 equinox}
\begin{tabular}{ccccc}
\hline
\hline
\multicolumn{1}{c}{Name} &
\multicolumn{1}{c}{$RA$}  &
\multicolumn{1}{c}{$DEC$}  &
\multicolumn{1}{c}{$l$} &
\multicolumn{1}{c}{$b$} \\
\hline
& {\rm $hh:mm:ss$} & {\rm $^{o}$~:~$^{\prime}$~:~$^{\prime\prime}$} & [deg] & [d
eg]\\
\hline
Berkeley~73        & 06:22:06 & -06:19:00 & 215.28 & -09.42\\
Berkeley~75        & 06:48:59 & -23:59:30 & 234.30 & -11.12\\
Berkeley~25        & 06:41:16 & -16:29:12 & 226.61 & -09.69\\
\hline\hline
\end{tabular}
\end{table}

   \begin{figure}
   \centering
   \includegraphics[width=9cm]{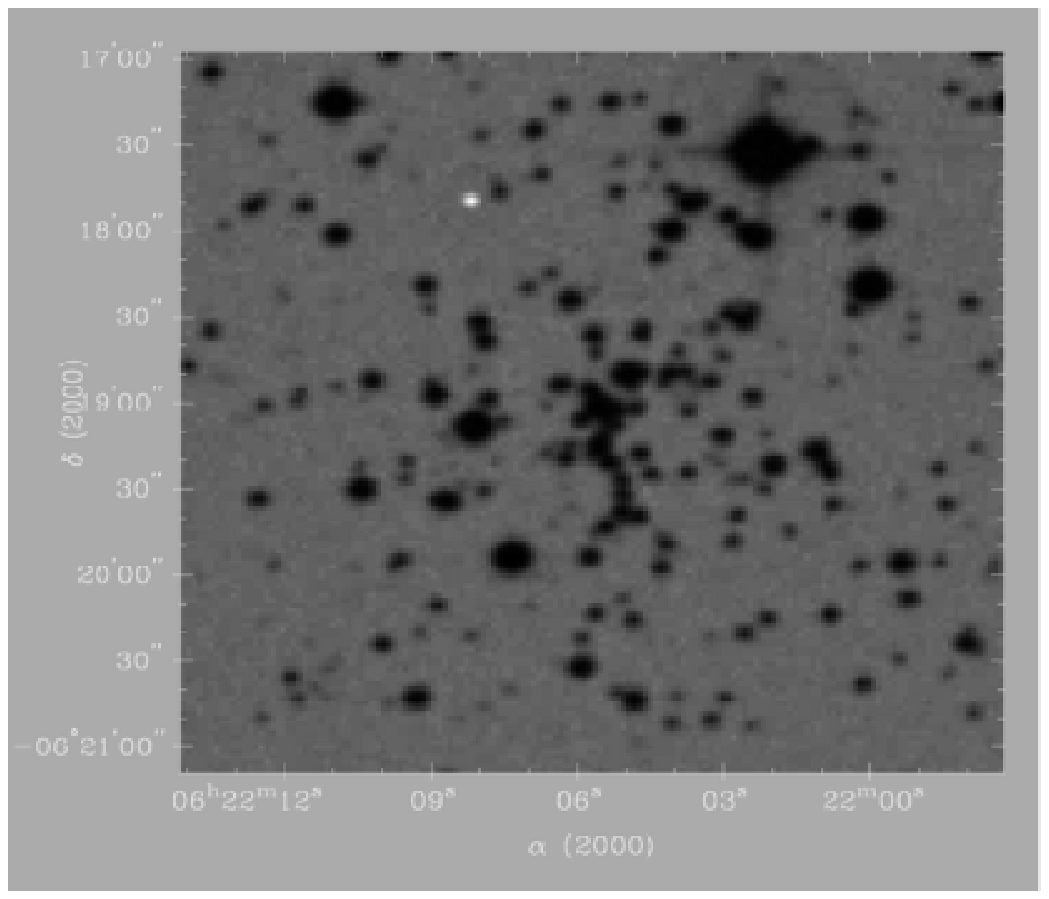} 
   \caption{A DSS red map of the covered region in the field
of Berkeley~73. North is up, East on the left.}
    \end{figure}

   \begin{figure}
   \centering
   \includegraphics[width=9cm]{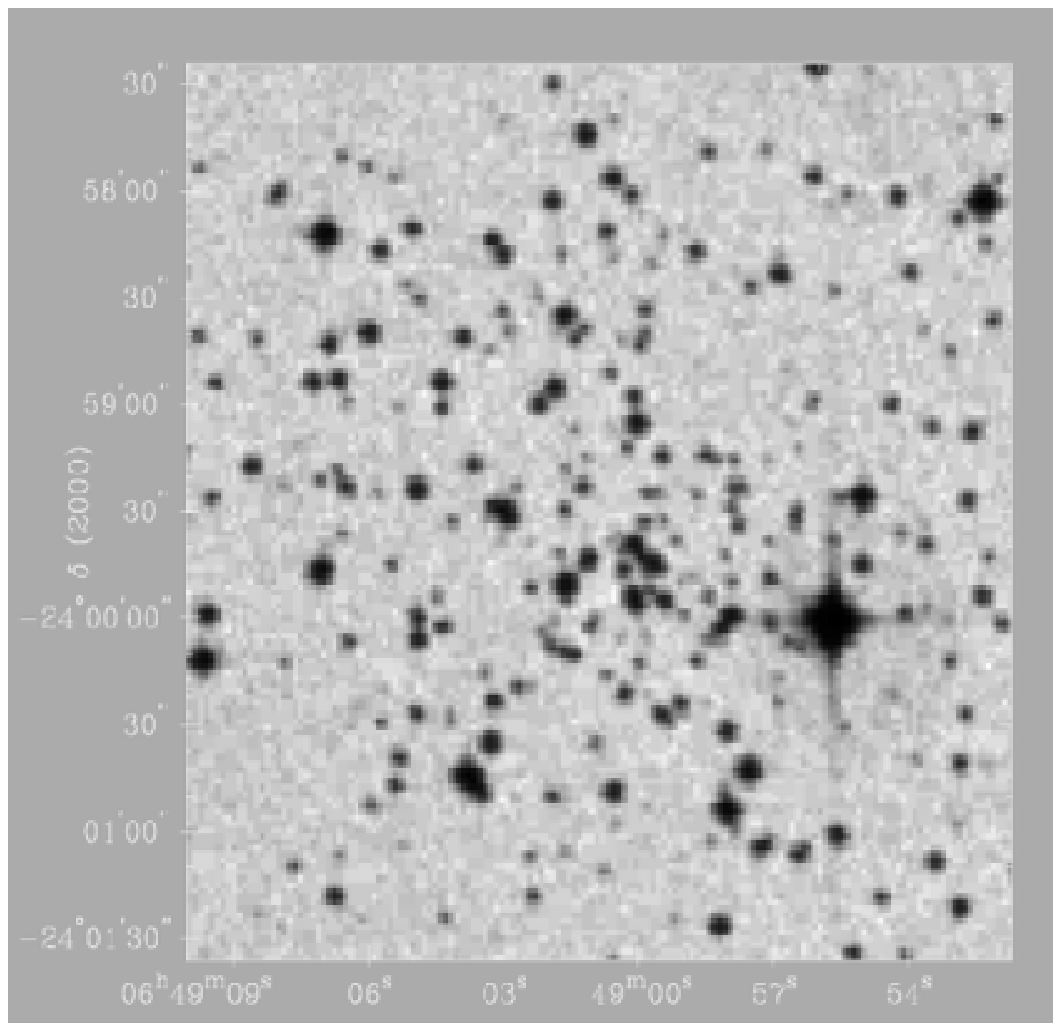} 
   \caption{A DSS red map of the covered region in the field
of Berkeley~75. North is up, East on the left.}
    \end{figure}

   \begin{figure}
   \centering
   \includegraphics[width=9cm]{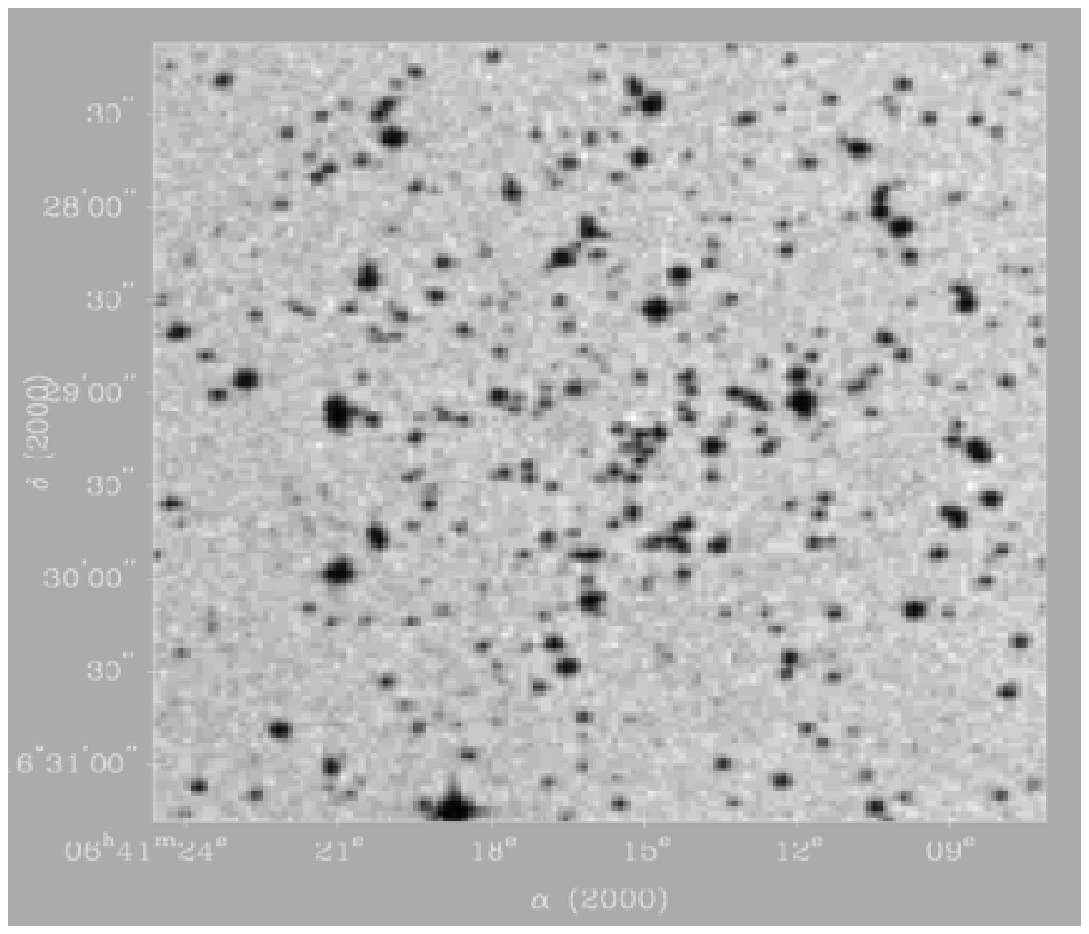} 
   \caption{A DSS red map of the covered region in the field
of Berkeley~25. North is up, East on the left.}
    \end{figure}

\section{Observations and Data Reduction}
CCD $BVI$ observations were carried out with the CCD camera on-board
the  1. 0m telescope at Cerro Tololo Interamerican Observatory (CTIO,Chile), on the nights of
December 13 and 15, 2004.
With a pixel size of $0^{\prime\prime}.469$,  and a CCD size of 512 $\times$ 512
pixels,
this samples a $4^\prime.1\times4^\prime.1$ field on the sky.\\
\noindent
The details of the observations are listed in Table~2 where the observed
fields are
reported together with the exposure times, the average seeing values and the
range of air-masses during the observations.
Figs.~1 to 3 show DSS finding charts in the area of
Berkeley~73, Berkeley~75 and Berkeley~25, respectively, and here
one can easily recognize that these clusters are faint
objects.

\noindent
The data have been reduced with the
IRAF\footnote{IRAF is distributed by NOAO, which are operated by AURA under
cooperative agreement with the NSF.}
packages CCDRED, DAOPHOT, ALLSTAR and PHOTCAL using the point spread function (PSF)
method (Stetson 1987).
The two nights turned out to be photometric and very stable, and therefore
we derived calibration equations for all the 130 standard stars
observed during the two nights in the Landolt
(1992)  fields SA~95-41, PG~0231+051, Rubin~149, Rubin~152,
T~phe and    SA~98-670 (see Table~2 for details).
Together with the clusters, we observed three control fields  20 arcmins
apart from the nominal cluster centers to deal with field star
contamination. Exposure of 600 secs in V and I were secured for these
fields.

\begin{table}
\fontsize{8} {10pt}\selectfont
\tabcolsep 0.10truecm
\caption{Journal of observations of Berkeley~73, Berkeley~75 and Berkeley~25
and standard star fields (December 13 and 15, 2004).}
\begin{tabular}{cccccc}
\hline
\multicolumn{1}{c}{Field}         &
\multicolumn{1}{c}{Filter}        &
\multicolumn{1}{c}{Exposure time} &
\multicolumn{1}{c}{Seeing}        &
\multicolumn{1}{c}{Airmass}       \\
 & & [sec.] & [$\prime\prime$] & \\
\hline
Berkeley~73    & B &       120,1200   &   1.2 & 1.12-1.20 \\
              & V &      30,600    &   1.3 & 1.12-1.20 \\
              & I &      30,600    &   1.2 & 1.12-1.20 \\
\hline
Berkeley~75    & B &         120,1200   &   1.2 & 1.12-1.20 \\
              & V &      30,600    &   1.3 & 1.12-1.20 \\
              & I &      30,600    &   1.2 & 1.12-1.20 \\
\hline
Berkeley~25     & B &    120,1200   &   1.2 & 1.12-1.20 \\
              & V &      30,600    &   1.3 & 1.12-1.20 \\
              & I &      30,600   &   1.2 & 1.12-1.20 \\
\hline
SA 98-670     & B &   $3 \times$120   &   1.2 & 1.24-1.26 \\
              & V &   $3 \times$40    &   1.4 & 1.24-1.26 \\
              & I &   $3 \times$20    &   1.4 & 1.24-1.26 \\
\hline
SA 95-041     & B &   $3 \times$120   &   1.2 & 1.24-1.26 \\
              & V &   $3 \times$40    &   1.4 & 1.24-1.26 \\
              & I &   $3 \times$20    &   1.4 & 1.24-1.26 \\
\hline
PG 0231+051   & B &   $3 \times$120   &   1.2 & 1.20-2.04 \\
              & V &   $3 \times$40    &   1.5 & 1.20-2.04 \\
              & I &   $3 \times$20    &   1.5 & 1.20-2.04 \\
\hline
T Phe         & B &   $3 \times$120   &   1.2 & 1.04-1.34 \\
              & V &   $3 \times$ 40   &   1.3 & 1.04-1.34 \\
              & I &   $3 \times$ 20   &   1.3 & 1.04-1.34 \\
\hline
Rubin 152     & B &   $3 \times$120   &   1.3 & 1.33-1.80 \\
              & V &   $3 \times$40    &   1.2 & 1.33-1.80 \\
              & I &   $3 \times$20    &   1.2 & 1.33-1.80 \\
\hline
Rubin 149     & B &   $3 \times$120   &   1.3 & 1.33-1.80 \\
              & V &   $3 \times$40    &   1.2 & 1.33-1.80 \\
              & I &   $3 \times$20    &   1.2 & 1.33-1.80 \\
\hline
\hline
\end{tabular}
\end{table}

\noindent
The calibration equations turned out of be of the form:\\

\noindent
$ b = B + b_1 + b_2 * X + b_3~(B-V)$ \\
$ v = V + v_1 + v_2 * X + v_3~(B-V)$ \\
$ v = V + v_{1,i} + v_{2,i} \times X + v_{3,i} \times (V-I)$ \\
$ i = I + i_1 + i_2 * X + i_3~(V-I)$ ,\\

\noindent

\begin{table}
\tabcolsep 0.3truecm
\caption {Coefficients of the calibration equations}
\begin{tabular}{ccc}
\hline
$b_1 = 3.465 \pm 0.009$ & $b_2 =  0.25 \pm 0.02$ & $b_3 = -0.145 \pm 0.008$ \\
$v_1 = 3.244 \pm 0.005$ & $v_2 =  0.16 \pm 0.02$ & $v_3 =  0.021 \pm 0.005$ \\
$v_{1,i} = 3.244 \pm 0.005$ & $v_{2,i} =  0.16 \pm 0.02$ & $v_{3,i} =  0.009 \pm0.005$ \\
$i_1 = 4.097 \pm 0.005$ & $i_2 =  0.08 \pm 0.02$ & $i_3 =  0.006 \pm 0.005$ \\
\hline
\end{tabular}
\end{table}

\noindent
where $BVI$ are standard magnitudes, $bvi$ are the instrumental ones and  $X$ is
the airmass; all the coefficient values are reported in Table~3.
The standard
stars in these fields provide a very good color coverage.
The final {\it r.m.s.} of the calibration are 0.039, 0.034 and 0.033 for the B, V and I filters,
respectively.
\noindent
We generally used the third equation to calibrate the $V$ magnitude
in order to get the same magnitude depth both in the cluster
and in the field.
It turns out that the limiting magnitudes are B = 21.9, V = 22.5
and I =21.8.
Moreover we performed a completeness analysis following the method described in
Carraro et al. (2005). It turns out that our sample 
has completeness level larger than 50$\%$ down to B = 20.0, V = 21.0
and I = 20.5.

The final photometric catalogs for 
Berkeley~73, Berkeley~75 and Berkeley~25 (coordinates,
B, V and I magnitudes and errors)
consist of 941, 1035 and 1039 stars, respectively, and are made
available in electronic form at the
WEBDA\footnote{http://obswww.unige.ch/webda/navigation.html} site
maintained by J.-C. Mermilliod.\\

\section{Astrometry}
We derived the astrometric solution to provide
for all the stars in the three clusters the J2000.0 coordinates.
This is a basic step for any further
follow-up study in the area of the cluster.
In order to obtain an astrometric solution we use the SkyCat tool
and the Guide Star Catalogue v2 (GSC-2) at ESO. This way we find about
80 stars per field for which we have both the celestial
coordinates on the GSC-2 and the corresponding pixel
coordinates. Then, by using the IRAF tasks CCXYMATCH, CCMAP and
CCTRAN, we find the corresponding transformations between the two
coordinate systems compute the individual celestial
coordinates for all the detected stars. The transformations have an
\textit{r.m.s.} value of $0\farcs37$ both in RA and in DEC, 
in broad agreement with other studies (Momany et al. 2001, Carraro et al. 2005).

\section{Star counts and cluster size}
Since our photometry covers entirely each cluster's area
we performed star counts to obtain
the first quantitative  estimate of the clusters size.
We derived the surface stellar density by performing star counts
in concentric rings around the clusters nominal centers (see Table~1)
and then dividing by their
respective area. Poisson errors have also been derived and normalized
to the corresponding area.
The field star contribution has been derived from the control
field which we secured for each cluster.
We selected the magnitude interval $12 \leq V \leq 20$ both in the cluster
and in the field to minimize incompleteness effects.\\

\noindent
{\bf Berkeley~73.}
The final radial density profile for Berkeley~73 is shown in 
the lower panel of Fig.~4. 
The profile is smooth and reaches the level of the field at
about 1.5 arcmin, which we are going to adopt as
Berkeley~73
radius throughout this paper. This estimate is a bit larger than
the value of 2.0 arcmin reported by Dias et al. (2002) for the cluster
diameter.\\

\begin{figure}
\centering
\includegraphics[width=8cm]{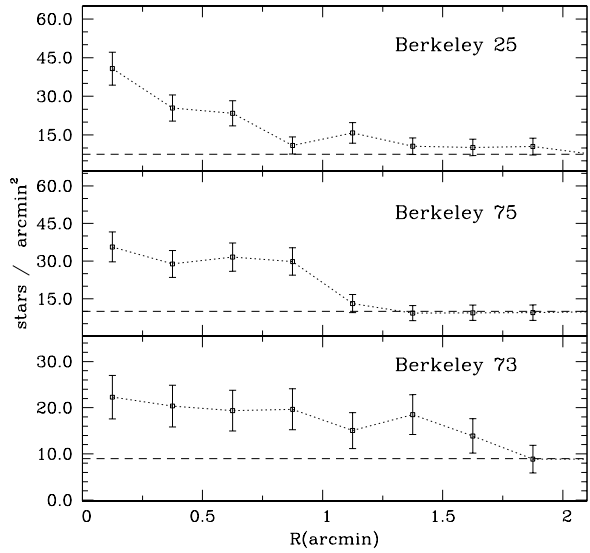}
\caption{Star counts in the area of the three open clusters
under study as a function of  radius. The dashed lines represent
the level of the control field counts estimated from the accompanying
control field.}
\end{figure}

\noindent
{\bf Berkeley~75.}
The final radial density profile for Berkeley~75 is shown in
the middle panel of  Fig.~4.
The cluster clearly emerges from the background up to
about 1.0 arcmin, a value that we  are going to adopt as
Berkeley~75
radius throughout this paper.\\

\noindent
{\bf Berkeley~25.}
The final radial density profile for Berkeley~25 is shown in the upper panel
of Fig.~4.
The cluster seems to be faint and dense, with a probable radius
of less than 1 arcmin.
This estimate is much smaller than
the value of 5.0 arcmin reported by Dias et al. (2002) for the cluster
diameter.\\

\noindent
The estimates we provide for the radius, although reasonable,
must be taken as preliminary.
In fact the size of the CCD is probably too small to derive
conclusive estimates of the cluster sizes. This is particularly
true in the case Berkeley~73, for which the cluster radius we derived
must be considered as a lower limit to the real cluster radius.
In fact, while for Berkeley 75 and 25 the cluster density profile
converges toward the field level within the region covered by the CCD,
in the case of Berkeley~73 the cluster dominates the star counts 
up to almost the border of the region we covered. Larger field coverage
is necessary in this case to derive a firm estimate of the cluster
radius.

\begin{figure*}
\centering
\includegraphics[width=19cm]{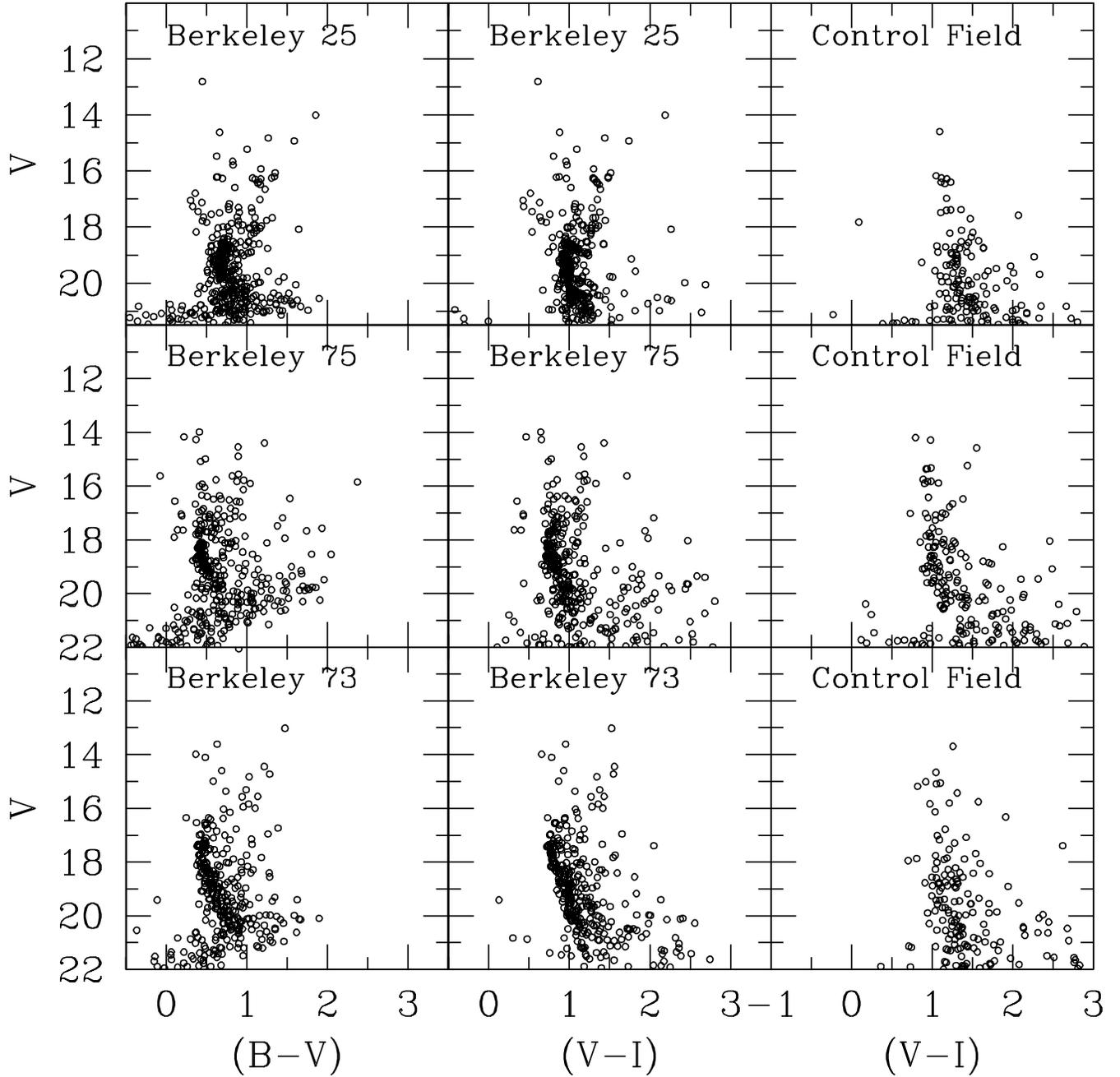}
\caption{$V$ vs $(B-V)$ (left panels) and $V$ vs $(V-I)$ (middle panels) CMDs
of Berkeley~73 (lower panels), Berkeley~75 (middle panels) and
Berkeley~25 (upper panels) and corresponding control fields (right panels).
We include all stars in each field.}
\end{figure*}

\begin{figure*}
\centering
\includegraphics[width=17cm]{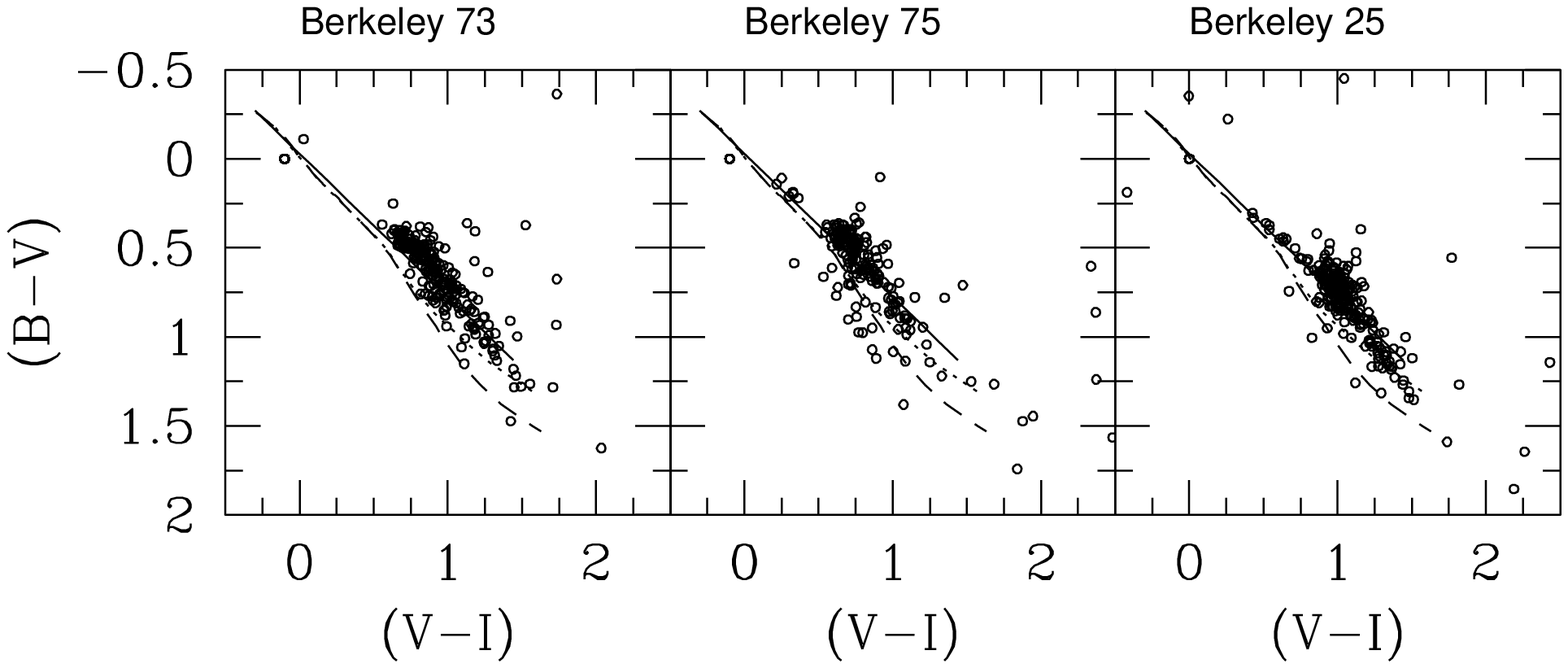}
\caption{$B-V$ vs $V-I$ diagram for the four clusters. The solid line
is the normal reddening law, whereas dashed and dotted lines are the
intrinsic colors for luminosity class III and V, respectively. See text for deta
ils.}
\end{figure*}

\section{Colour-Magnitude Diagrams}

In Fig.~5 we present the CMDs we obtained for the three clusters
under investigation.
In this figure the open cluster Berkeley~73 is shown  together
with the corresponding control field in the lower panels,
whereas Berkeley~75 and Berkeley~25 are presented in the
middle and upper panels, respectively. The control fields
help us to better interpret these CMDs, which are clearly
dominated by foreground star contamination.\\

\noindent
{\bf Berkeley~73}.
It exhibits a Main Sequence (MS) extending from
V=17, where the Turn Off Point (TO) is located, down
to V=21. This MS is significantly wide, a fact that we
ascribe to the increasing photometric error at increasing magnitude,
the field star contamination,
and to the possible presence of a sizeable binary star population,
which mainly enlarges the MS toward red colors.
However, the reality of this cluster seems to be secured by the shape and
density of
the MS compared  to the control field MS, whose
population sharply decreases at V $\leq$ 18.
In addition, the cluster MS is significantly bluer and more tilted
than the field MS, which derives from the superposition
of stars of different reddening located at all distances.
Another interesting evidence it the possible presence of  a clump
of stars at V=15.5-16.0, which does not have a clear counterpart
in the field, and which makes the cluster an intermediate-age one.
In fact if we use the age calibration from Carraro \& Chiosi (2004),
for a $\Delta V$ (the magnitude difference between
the red clump and the TO) of 1.5 mag, we infer an age around 1.2 billion
year. This estimate does not take into account the cluster metallicity,
and therefore is a simple guess. In the following we shall
provide a more robust estimate of the age through a detailed
comparison with theoretical isochrones.\\

\noindent
{\bf Berkeley~75}.
The TO is located at $V \approx$ 18, and the clump
at $V \approx$16.0, thus implying a rough estimate for the age around 3 billion
year. The overall morphology of the CMDs is less different from the
field  than the previous cluster.
A significant binary population
seems to affect the TO shape, which appears at first glance
confused (see for comparison the CMDs in the middle
and upper panels). \\
Moreover, the field star contamination seems to be
 important for this cluster, but it does not prevent us
from recognizing a probable clump
of stars  located at V =  16, B-V = 1.2, (V-I)= 1.4.
The field in the right panel also exhibits a clump, but it lies
along the MS at (V-I) = 1.0.
If the Berkeley~75 clump mean magnitude is actually 
around V=16,  
this would imply a 
distance modulus (m-M) of about 16 mag, according
to Salaris \& Girardi (2002).\\

\noindent
{\bf Berkeley~25}.
The TO is located at $V \approx$ 18.5, and the clump
at $V \approx$ 16.5, thus implying a rough estimate for the age 
around 3.0 Gyr
year. The overall morphology of the CMDs is also in this case
very different from the
field CMD so this is a bona-fide cluster.
The same kind of argument applied to Berkeley~75 clump 
holds for Berkeley~25, since also in this case the field exhibits
a star condensation at about the cluster
clump position.

\section{Clusters fundamental parameters}
In this section we perform a detailed comparison
of the star distribution in the clusters CMDs with theoretical
isochrones. We adopt in this study the Padova library from
Girardi et al. (2000).
This comparison is clearly not an easy exercise. In fact
the detailed shape and position of the various features
in the CMD (MS, TO and clump basically) depends mostly
on age and metallicity, and then also on reddening and distance.
The complex interplay between the various parameters is well
known, and we refer to
Chiosi et al. (1992) and Carraro (2005)
as nice examples of the underlying technique.\\
Our basic strategy is to survey different age and metallicity
isochrones attempting to provide the best fit of all the CMD
features both in the $V$ vs $(B-V)$ and in the $V$ vs $(V-I)$ CMD.\\
Besides, to further facilitate the fitting procedure
we shall consider only the stars which lie  within
the cluster radius as derived in Sect.~3.
\noindent
Therefore, in the series of Figs.~7 to 9 we shall present the best fit
we were able to achieve.
Together with the best fit, we could make estimates of
uncertainties in the basic parameters derivation. These uncertainties simply
reflect the range in the basic parameter
which allow a reasonable fit to the clusters CMDs. In particular in the
left panel of  Figs.~7 to 9, we  overlaid another isochrone to show
the effect of assuming different combinations of age and metallicity.
Error estimates are reported in Table~4.\\

\noindent
To derive clusters' distances from reddening and apparent distance modulus,
a reddening law must be specified. In Fig.~6 we show that the normal extinction
law is valid for all the clusters, and therefore we shall us the relation
$Av = 3.1 \times E(B-V)$ to derive clusters' distances. 
In details, the solid line
in Fig.~10 is the normal extinction $E(V-I)=1.245 \times E(B-V)$ law, whereas th
e dashed and dotted lines are
the luminosity class $III$ and $V$ ZAMS from Cousin (1978).
Finally to limit in some ways the degrees of freedom of the whole 
fit technique, we derived from FIRB maps
(Schlegel et al. 1998) the reddening in the direction of the clusters.
We got E(B-V) = 0.18, 0.12 and 0.10 for Berkeley~73, Berkeley~75 and Berkeley~25,
respectively. Due to the clusters Galactic latitude, these values
can be considered very reasonable.\\

\noindent
{\bf Berkeley~73}.
The isochrone solution for this cluster is discussed in Fig.~7.
We consider here only the stars located within 1.5 arcmin from the cluster
center (see Sect.~2).
We obtained the best fit for an age of 1.5 Gyr and a metallicity
Z=0.008. The inferred
reddening and apparent distance modulus are E(B-V)=0.12 (E(V-I)=0.18)
and (m-M)=15.3, respectively. As a consequence the cluster
possesses a heliocentric distance of 9.7 kpc, and
is located at a Galactocentric distance of 16.4  kpc, assuming
8.5 kpc as the distance of the Sun to the Galactic Center.
The overall fit is very good, the detailed shape of the MS
and TO are nicely reproduced, and the color of the clump as well.
With respect to Ortolani et al. (2005) we find a significantly
younger age, and this is due to our interpretation of the
group of stars at V $\approx$ 16, (B-V) $\approx$ 1.0 as a clump He-burning
stars. This choice is supported by the fact that we are selecting only stars
located well within the cluster radius, and actually this group of stars
lie very close to the cluster center, as expected for more massive stars
after the mass segregation mechanism.
Looking carefully at the isochrone fitting in Ortolani et al. (2005, Fig.~9),
one can note that the isochrone poorly matches the low MS, below V $\approx$ 19,
and is in general too red with respect to the mean locus of Berkeley~73 stars.
Also the color of the TO point is poorly matched. To achieve a better fitting
a bluer (say younger) isochrone is necessary.\\
To address the effect of a possible variation of the clusters parameters due to a different
metallicity, in the left panel of Fig.~7 we superposed the best fit Z=0.019
isochrone, which provides an age of 2 Gyr and a distance modulus 
(m-M)=14.8. The fit is rather good in the MS, but the RGB is too blue,
like in Ortolani et al (2005). A lower metallicity isochrone would
solve this problem.
Moreover the reddening value we get in this case is
E(B-V)=0.00$\pm$0.03. This value of the reddening is reasonable indeed 
for this combination of age and metallicity.
If the Ortolani et al. (2005) fit matched the TO color, it
would imply exactly such a reddening, and not the value they report.\\
Therefore we favor a low metal content and a younger age (see Table~4).
Moreover the set of Galactic coordinates that Ortolani et al. (2005)
report are in error, due to the wrong Galactic $l$ and $b$ 
they assign to this cluster. Therefore the cluster does not
lie in the fourth Galactic quadrant, but below the plane in the third 
quadrant.\\
\noindent
Interestingly, a considerable binary fraction seems to be present in the
form of a parallel sequence red-ward the cluster MS.\\

\begin{figure}
\centering
\includegraphics[width=9cm]{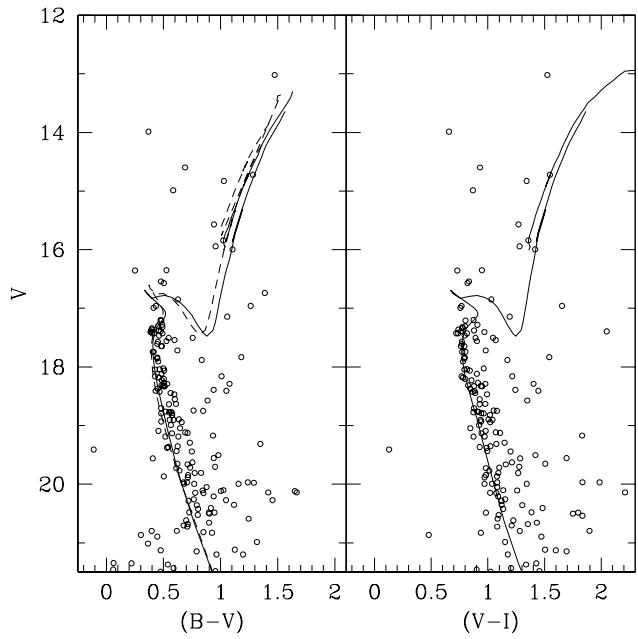}
\caption{Isochrone solution for Berkeley~73. The solid lines are isochrones 
for the age of 1500 million year and metallicity Z=0.008.
The apparent distance modulus is (m-M)=15.3, and the reddening
E(B-V)=0.12 and E(V-I)=0.18. 
Only stars within the derived radius are shown. The dashed line in the left
panel is a Z=0.019 isochrone for the age of 2 Gyr. See text for more details.}
\end{figure}

\noindent
{\bf Berkeley~75}
The isochrone solution for this cluster is discussed in Fig.~8.
We obtained the best fit for an age of 3 Gyr and a metallicity
Z=0.004. 
The inferred
reddening and apparent distance modulus are E(B-V)=0.08 (E(V-I)=0.13)
and (m-M)=15.2, respectively. As a consequence the cluster
lies at 9.8 kpc from the Sun, and
is located at a Galactocentric distance of 16.2 kpc
toward the anti-center direction in the third Galactic quadrant.
The overall fit is very good also in this case, the detailed shape of the MS
and TO are nicely reproduced, and the color of the clump as well.\\
The reddening we derived by the way is in close agreement with FIRB maps,
which at the Galactic latitude of Berkeley~75 are surely reliable.\\
Also in this case we check the age-metallicity effect. Superposed in the
left panel of Fig.~8 is a Z=0.008 isochrone for the age of 3.5 Gyr, which
produces the best fit to the bulk of Berkeley~75 stars.
By assuming this combination of age and metallicity, the derived
distance modulus is (m-M)=14.9, and the reddening E(B-V) = 0.00.
By increasing more the metallicity an untenable negative reddening would
be necessary to get a good fit.
We therefore conclude that the best fit metallicity is Z=0.004,
which yields the parameters set reported in Table~4.\\

\begin{figure}
\centering
\includegraphics[width=9cm]{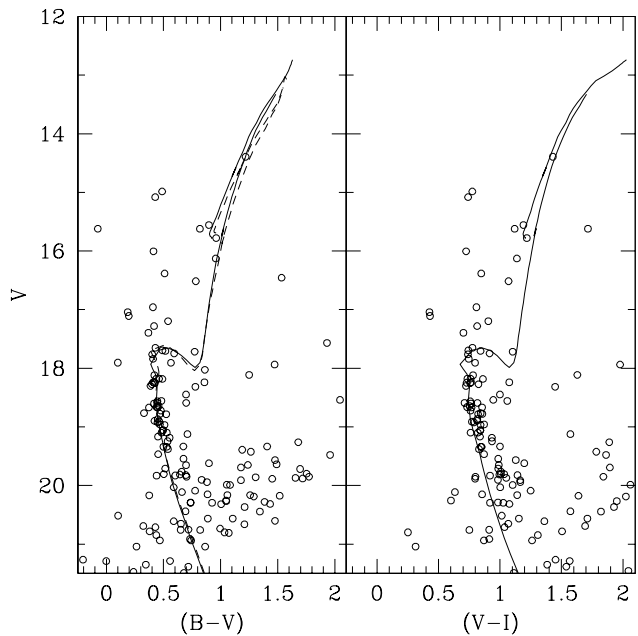}
\caption{Isochrone solution for Berkeley~75. The isochrones (solid lines)
are for the age of 3 Gyr and metallicity Z=0.004.
The apparent distance modulus is (m-M)=15.2, and the reddening
E(B-V)=0.08 and E(V-I)=0.13. 
Only stars within the derived radius are shown. 
The dashed line is a 3.5 Gyr, Z=0.008 isochrone.
See text for more details.}
\end{figure}

\noindent
{\bf Berkeley~25}
The isochrone solution for this cluster is shown in Fig.9.
We obtained the best fit for an age of 4 Gyr and a metallicity
Z=0.008. The inferred
reddening and apparent distance modulus are E(B-V)=0.17 (E(V-I)=0.24)
and (m-M)=15.8, respectively. Therefore the cluster has a heliocentric distance
of 11.3 kpc,
and
is located at a Galactocentric distance of 16.8 kpc.
The overall fit is very good also in this case, the detailed shape of the MS
and TO are nicely reproduced, and the color of the clump as well.
The reddening we derived by the way is in close agreement with FIRB maps.\\
In the left panel of the same figure we overlaid a metal poorer (Z=0.004) 
isochrone for the age of 3.5 Gyrs. This isochrone provides a good fit
as well, with a distance modulus (m-M)=16.2, and a reddening E(B-V)=0.35.
However, since FIRB maps support a much lower reddening, we tend to favour
here the higher metallicity solution, which provides a more reasonable reddening
value.\\

\begin{figure}
\centering
\includegraphics[width=9cm]{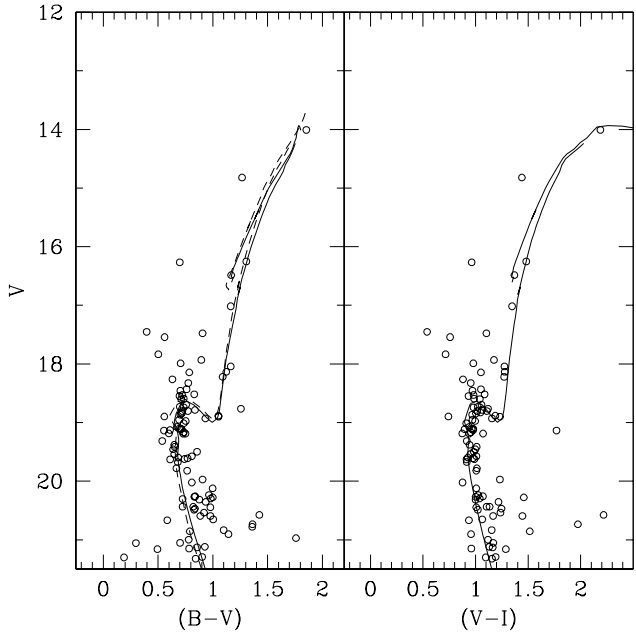}
\caption{Isochrone solution for Berkeley~25. The isochrones (solid lines)
are for the age of 4 Gyrs year and metallicity Z=0.008.
The apparent distance modulus is (m-M)=15.8, and the reddening
E(B-V)=0.17 and E(V-I)=0.24. 
Only stars within the derived radius are shown.
The dashed line is a 3.5 Gyr, Z=0.004 isochrone.
See text for more details.}
\end{figure}

\noindent
All these clusters are very interesting because of their position so far
away from the Galactic center. We only known in fact a few clusters
located so distant from the center of the Galaxy (Villanova et al. 2005),
and therefore they are crucial targets for spectroscopic follow-up.
Besides, Berkeley~25 is one of the old open clusters located at highest Galactic 
latitude (see Table~4, and Friel 1995).\\

\begin{table*}
\caption{Fundamental parameters of the studied clusters. The coordinates system
is such that
the Y axis connects the Sun to the Galactic Center, while the X axis is perpendicular 
to that.
Y is positive toward the Galactic anti-center, and X is positive in the first and
 second Galactic quadrants (Lynga 1982).}
\fontsize{8} {10pt}\selectfont
\begin{tabular}{cccccccccccc}
\hline
\multicolumn{1}{c} {$Name$} &
\multicolumn{1}{c} {$Radius$} &
\multicolumn{1}{c} {$E(B-V)$}  &
\multicolumn{1}{c} {$E(V-I)$}  &
\multicolumn{1}{c} {$(m-M)$} &
\multicolumn{1}{c} {$d_{\odot}$} &
\multicolumn{1}{c} {$X_{\odot}$} &
\multicolumn{1}{c} {$Y_{\odot}$} &
\multicolumn{1}{c} {$Z_{\odot}$} &
\multicolumn{1}{c} {$R_{GC}$} &
\multicolumn{1}{c} {$Age$} &
\multicolumn{1}{c} {Metallicity}\\
\hline
& $^{\prime}$ & mag & mag & mag& kpc & kpc & kpc & kpc & kpc & Gyr & \\
\hline
Berkeley~73 &  1.5 & 0.12$\pm$0.05 & 0.18$\pm$0.05  & 15.3$\pm$0.2 & 9.7 & -5.0 &  7.1 & -1.40 & 16.4& 1.5$\pm$0.2 & 0.008$\pm$0.004\\
Berkeley~75 &  1.0 & 0.08$\pm$0.05 & 0.13$\pm$0.05  & 15.2$\pm$0.2 & 9.8 & -7.8 &  5.6 & -1.90 & 16.2& 3.0$\pm$0.3 & 0.004$\pm$0.002\\
Berkeley~25 &  0.8 & 0.17$\pm$0.05 & 0.24$\pm$0.05  & 15.8$\pm$0.5 &11.3 & -9.9 &  4.9 & -1.90 & 16.8& 4.0$\pm$0.5 & 0.008$\pm$0.003\\
\hline
\end{tabular}
\end{table*}

\section{Conclusions}
We have presented CCD $BVI$ photometric study of the
star clusters Berkeley~73, Berkeley~75 and Berkeley~25.
The CMDs we derive allow us to
infer estimates of the clusters' basic parameters, which
are summarized in Table~4.\\
\noindent
In detail, the fundamental findings of this paper are:

\begin{description}
\item $\bullet$ the best fit reddening estimates support
within the errors a normal extinction law toward the three clusters;
\item $\bullet$ all the clusters are very distant from the Galactic center
and quite high onto the Galactic plane.
They might be very interesting object for further investigation
to clarify whether they might belong to the recently discovered
Canis Major-Monoceros  over-density;
\item $\bullet$ Berkeley~73 is found to be of intermediate age.
\end{description}

\noindent
All these clusters are very interesting in the context of the
chemical evolution of the Galactic disk
(Geisler et al. 1992, Carraro et al. 1998, Friel et al 2002).\\
They may be very useful
to trace the slope of the Galactic disk radial abundance gradient
and to probe the chemical properties and the structure of the disk in its outskirts,
in a region which is dominated by the Galactic warp and/or the recently discovered
Canis Major over-density.\\
If we make use of the provisional photometric estimates
presented in this paper, we obtain that all the clusters 
are basically consistent
with the most recent slope of the gradient in the  3-4 Gyrs and  $\leq$ 2 
Gyr bins, as derived by Friel et al. (2002). \\
\noindent
Further studies therefore should concentrate on the confirmation of the clusters'
metal content by means of a detailed abundance analysis of the
clump stars.\\

\noindent
Finally, Berkeley~75 and Berkeley~25 are very interesting clusters
because they fall in an age range where only a few clusters are known
(see the discussion in Ortolani et al. 2005). 
The authors show the age distribution of 103 old open clusters.
The ages they use are highly inhomogeneous, since they come from a variety
of different sources. This fact however does not seem to change 
significantly the overall conclusions
on the possible existence of a peak at 5 Gyrs with respect to
the previous analysis on the same subject conducted by Friel (1995)
using a smaller but homogeneous sample.
This might mean that the level of inhomogeneity in the ages is not very important.\\
\noindent
 If we include the new two 
clusters Berkeley~75 and Berkeley~25,  Berkeley~22 and Berkeley~66 (around 3-4 Gyrs old, 
Villanova et al. 2005), and Berkeley~29 (4 Gyrs, Carraro et al. 2004),
for which we report new estimates of the age, 
the dip Ortolani et al. (2005) report in the old open clusters age distribution
(their Fig.~11) becomes less prominent, and the reality of a
peak at 5 Gyrs less significant. The new age distribution is shown
in Fig.~10, where we consider all the clusters reported by Ortolani et al.
(2005, solid line) and add a few more clusters,
(dashed line).\\
It is by the way quite probable
that more clusters are to be discovered in this age range.\\

\noindent
In our opinion, this is simply suggesting that the completeness of the sample  
and homogeneity of the ages are  crucial issues which have to be carefully
taken into account before drawing conclusions 
on the star formation history of the Galactic disk
using old open clusters.

\begin{acknowledgements}
We thank the anonymous referee for important suggestions,
that led to a better presentation of the paper.
The observations presented in this paper have been carried out at
Cerro Tololo Interamerican Observatory CTIO (Chile).
CTIO is operated by the Association of Universities for Research in Astronomy,
Inc. (AURA), under a cooperative agreement with the National Science Foundation
as part of the National Optical Astronomy Observatory (NOAO).
The work of G.C. is supported by {\it Fundaci\'on Andes}.
D.G. gratefully acknowledges support from the Chilean
{\sl Centro de Astrof\'\i sica} FONDAP No. 15010003.
This work has been also developed in the framework of
the {\it Programa Cient\'ifico-Tecnol\'ogico Argentino-Italiano SECYT-MAE
C\'odigo: IT/PA03 - UIII/077 - per\'iodo 2004-2005}.
A.M. acknowledges support from FCT (Portugal) through grant
SFRH/BPD/19105/2004.
This study made use of Simbad and WEBDA databases.

\begin{figure}
\centering
\includegraphics[width=9cm]{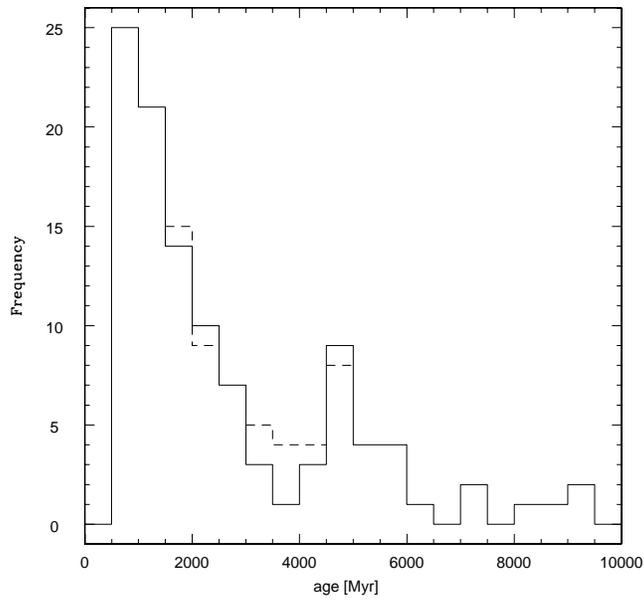}
\caption{Age histogram of old open clusters. See text for details}
\end{figure}

\end{acknowledgements}

\end{document}